# Multiphysics Modeling on Photoconductive Antennas for Terahertz Applications

Boxun Yan[1], Bundel Pooja[2], Chi-Hou Chan[2], and Mau-Chung Frank Chang[1]

[1]Department of Electrical and Computer Engineering, University of California, Los Angeles (UCLA),

Los Angeles, CA 90095, USA, boxun@ucla.edu

[2]State Key Laboratory of Terahertz and Millimeter Waves, City University of Hong Kong, HKSAR

*Abstract*—Terahertz lies at the juncture between RF and optical electromagnetism, serving as a transition from mm-Wave to infrared photonics. Terahertz technology has been used for industrial quality control, security imaging, and high-speed communications, and often generated through optoelectronic solutions by using photoconductive antennas. In this paper, Multiphysics simulations on semi-insulating GaAs, graphene-enhanced photoconductive antennas are conducted to effectively decouple optical responses of semiconductor carrier generation/drift from Terahertz radiation computation, which provides a comprehensive and integrated platform for future terahertz photoconductive antenna designs.

## I. Introduction

Terahertz (THz) waves occupy a unique position in the electromagnetic spectrum, bridging the gap between the conventional microwave/mmWave and photonics, thereby encapsulating dual natures of radio frequency (RF) and optics.

From a semiconductor device perspective, operating frequencies within RF bands encounter limitations from the finite time required for carriers to transit through control electrodes (either gate or base). On the other hand, operating frequencies within optical bands encounters limitations from the quantum levels constrained by the natural bandgaps of formation semiconductors.

These constraints were partially addressed by using Photoconductive Antennas (PCA) to leverage the rapid movement of electrons for THz generation [1] and extensive research was conducted on PCA design and simulations [2]. However, these computational solvers were Maxwell-equation based, often exhibiting a limited integration capacity, consequently falling short in comprehensiveness regarding THz generation mechanisms. Nonetheless, previous Multiphysics models were either limited to a 2D computational domain or ineffective for THz radiation computation [3, 4, 5].

In this paper, we handle the complexities of Multiphysics and multi-scale modeling, ranging from incident laser, carrier generation and drift within the semiconductor substrate to THz antenna modeling. Our numerical method is grounded in both electromagnetism and quantum mechanics and aims to bridge gaps between analytical understanding and practical application for PCAs.

Additionally, graphene is identified to provide performance enhancement for certain frequency ranges [6, 7]. By tying a sheet of graphene to the surface of a semi-insulating (SI) GaAs substrate, photo-generated carriers are transported to electrodes at a faster rate, revealed in Section IV.

The remainder of the paper is structured as follows: Section II presents the Multiphysics and Multi-scale modelling method. Section III and IV show the Multiphysics simulation results and analysis. Finally, summaries are given in Section V.

## II. Multi-physical Simulation

The Multiphysics model developed for PCA simulations is designed to capture the multifaceted nature of THz generation. It is structured to include multiple components, including the modeling for incident laser, which characterizes the initial electromagnetic energy input; the optical response of the SI-GaAs substrate; and carrier dynamics within the semiconductor, addressing the photo-generated carrier movements and interactions.

### A. Laser Modelling

The infrared laser is modelled in the frequency domain, with wavelength normally distributed by $\lambda \sim \mathcal{N}(\mu = 780, \sigma^2 = 10^2)$ in nm. Assuming the Gaussian beam has a spatial profile of $G_{env}(x, y)$, the modulated laser pulse is given in (1).

$$\mathbf{E}(z,t) = \hat{\mathbf{x}} E_m \cos(\omega t + \beta z) \, \exp\left(-N\frac{(t-t_0)^2}{\tau_l^2}\right) G_{env} \quad (1)$$

The general solution to an incident wave in a specific material with $\epsilon_r$ and $\mu_r$ in frequency domain is given by (2).

$$\nabla \times \mu_r^{-1}(\nabla \times \mathbf{E}) - k_0^2 \left(\epsilon_r - \frac{j\lambda\sigma}{2\pi c \epsilon_0}\right)\mathbf{E} = 0 \quad (2)$$

### B. Optoelectronic response

In response to the incident laser pulse, stimulated emission from the SI-GaAs substrate can be calculated by perturbing the system Hamiltonian under laser illumination. With a practical system whose polarization density is significantly larger than the charge potential, i.e., $q\mathbf{A} \ll \mathbf{p}$, we can ignore the term which contains higher orders of $q\mathbf{A}$. Thus, the Hamiltonian variation due to a single electron motion is given by (3).

$$H = \frac{1}{2m_0}(\mathbf{p} - q\mathbf{A})^2 + V \quad (3)$$

In (3), $\mathbf{A}$ is the magnetic vector potential. Its divergence is set to 0 for simplicity. With the approximation above, the total energy can be represented by (4).

$$H = \frac{1}{2m_0}p^2 + V + \frac{e}{2m_0}\mathbf{A} \cdot \mathbf{p} \quad (4)$$

By Fermi's Golden rule, the probability of transition per unit time, $W_{12}$ from state 1 to 2, is given by (5).

$$W_{12} = \frac{2\pi}{\hbar}|H'_{12}|^2 \delta(E_2 - E_1 - \hbar\omega_0) \quad (5)$$

The incident laser wavelength is typically much larger than the unit lattice size of the substrate. It is assumed to have a parabolic relation at $k = 0$, $E_{1v} = E_v - \frac{\hbar k_0^2}{2m_h}$; $E_{2c} = E_v + h\nu_0 = E_{1v} + \hbar\omega_0$. At the lowest energy, the bandgap is $E_c - E_v = E_g$. For a state given $k$ and a given excitation, an average matrix element $H_{12}^{av}$ is defined to represent the averaged matrix element in all directions. With the assumptions above, (5) now takes the form of

$$W_{12} = \frac{2\pi}{\hbar}|H_{12}^{av}|^2 \delta(E_2 - E_1 - \hbar\omega_0) \quad (6)$$

The stimulated absorption is thus given by (7).

$$G_{stim}^{12} = \iiint \frac{2\pi}{\hbar}|H_{12}^{av}|^2 g(k) f_v (1 - f_c) \delta(E_{2c} - E_{1v} - \hbar\omega_0) d^3k \quad (7)$$

In (7), $f_c$ and $f_v$ are occupation factors in the conduction and valance bands, respectively. The reduced density of states $g_{red}(E)$ given by (8).

$$g_{red}(E) = \frac{1}{2\pi^2}\left(\frac{2m_r}{\hbar^2}\right)^{\frac{3}{2}}\sqrt{(E - E_g)} \quad (8)$$

With the same arguments, the stimulated emission is given by (9).

$$R_{stim}^{21} = \frac{\hbar \pi^2 c^3 n_E}{n^3 \omega_0^2 \tau_{snn}} f_e (1 - f_v) g_{red}(\hbar\omega_0) \quad (9)$$

*C. Optical Properties*

Starting from Poynting's theorem, the power flux across a surface in a material is given by (10).

$$\int_S \mathbf{E} \times \mathbf{H} \cdot \mathbf{n} dS = \int_V \mathbf{E} \cdot \mathbf{J} + \frac{\partial}{\partial t}\left(\frac{\varepsilon_0}{2}|\mathbf{E}|^2 + \frac{\mu_0}{2}|\mathbf{H}|^2\right) + \mathbf{E} \cdot \frac{\partial \mathbf{P}}{\partial t} + \mu_0 \mathbf{H} \cdot \frac{\partial \mathbf{M}}{\partial t} dV \quad (10)$$

where the polarization $\mathbf{P} = \epsilon_0 \chi \mathbf{E}$ is the magnetization, and $\chi = \chi' - j\chi''$ is the material susceptibility. Consequently, the time-averaged power dissipated is given by (11).

$$\langle P_v \rangle = \frac{1}{2}\text{Re}[j\omega_0 E^*(r) \cdot P(r)] = \frac{2\hbar G_{stim}}{\epsilon_0 E_0^2} \quad (11)$$

The imaginary part of susceptibility is directly related to the illumination, computed in (12).

$$\chi''(\omega_0) = \frac{2\hbar G_{stim}}{\varepsilon_0 E_0^2} = \frac{4\pi}{\varepsilon_0 E_0^2}|H_{12}^{av}|^2 (f_v^0 - f_c^0) g_{red}^0 \quad (12)$$

In the simulation, first-order polynomial approximation is leveraged to estimate the change in optical properties.

*D. Metal-Semiconductor Contact and Graphene modelling*

An ohmic contact is assumed to model the AuGe-GaAs junction. The effective Richardson constants for electrons and holes are $A_n^* = 110$ and $A_p^* = 90 \text{ A/m}^2\text{K}^2$.

At the interface, the carriers will be absorbed by the electrode, and the terminal current is an integral of current density across the surface, given by

$$i = \int_S \mathbf{J} \cdot d\mathbf{A} \quad (13)$$

In (13), $\mathbf{J}$ is the current density inside the substrate induced by photo-generated carriers.

When tying a sheet of graphene atop the surface of SI-GaAs, the current density boundary condition can be specified as $\mathbf{J} = \sigma \mathbf{E}$. Its conductivity is calculated using Drude and Kubo's model [8].

*E. Optimization and Multiphysics coupling*

In computational EM, there are generally accepted restrictions on meshing. Conventionally, the dimension of maximum element rarely exceeds $\frac{\lambda}{6}$. However, the antenna is designed for THz radiation ($\lambda_{THz} \sim 1\text{mm}$) while laser has significantly smaller wavelengths ($\lambda_{laser} \sim 1\mu\text{m}$). This issue can be remedied by importing optoelectronic responses to EM without solving them simultaneously. The semiconductor model of the SI-GaAs model is highly nonlinear, and thus the solver sometimes encounters non-convergence problems. Semiconductor initialization is used to initialize the grid, and semiconductor stabilization yields a better initial value for convergence. Semiconductor and optics are coupled at the optoelectronic response and optical transitions stage. The Multiphysics coupling is shown in Fig. 1. Transient and frequency domain simulations are performed sequentially.

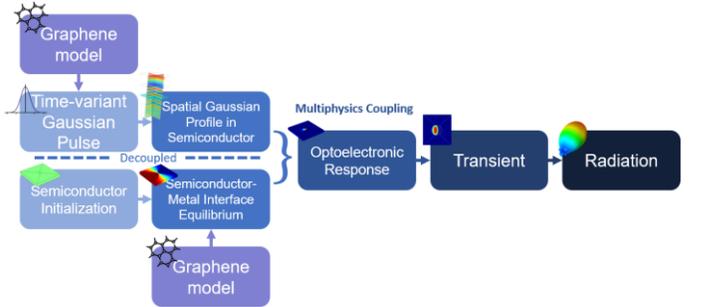

Fig. 1. Proposed Multiphysics simulation process

The multi-scale simulation is achieved by decoupling the optical response from THz radiation. The transient surface current density $\mathbf{J}_s$ on PCA is imported to EM simulation with a different mesh configuration. The radiated THz can be computed in (14).

$$\mathbf{E}_{THz} = \frac{1}{4\pi\epsilon c^2} \int \frac{\mathbf{J}_s\left(\mathbf{r}', t - \frac{|\mathbf{r}' - \mathbf{r}|}{c}\right)}{|\mathbf{r}' - \mathbf{r}|} ds' \quad (14)$$

Thus, a Multiphysics and multi-scale model is developed from optics laser, through a quantum process, to THz radiation.

## III. SIMULATION RESULTS

The transient simulation reveals the carrier dynamics in the SI-GaAs substrate. The simulation domain and transient electron density is plotted in Fig 2.

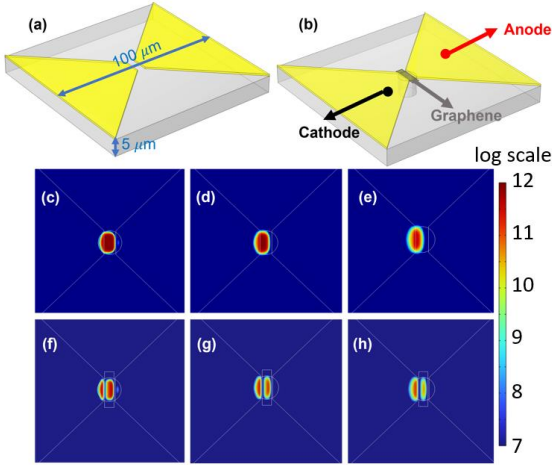

Fig. 2. (a-b) Computational domain; (c-e) electron density in log scale without graphene; (f-h) electron density in log scale with graphene, at t=1, 5, 10ps, respectively.

To study the transient current, an integral of the current density given by (13) is computed. The associated currents are shown in Fig. 3.

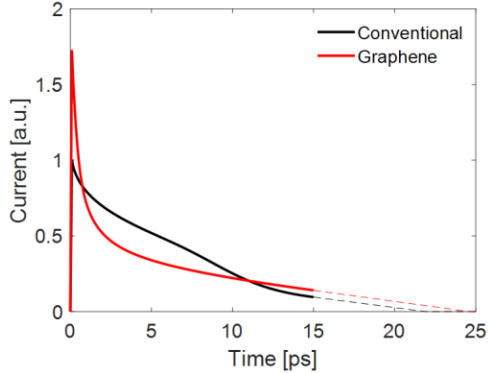

Fig. 3. Transient current on electrodes. The dashed line is extrapolated.

The simulated radiation intensity is shown in Fig. 4.

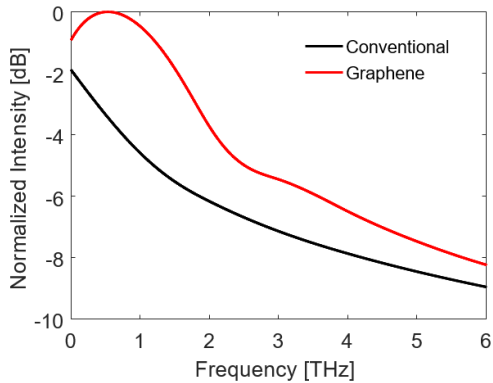

Fig. 4. Simulated radiation intensity.

## IV. ANALYSIS

The simulation shows that radiated THz waves are stimulated from the semiconductor by incident femtosecond laser pulse. The terminal current acts as the stimulus for the radiating antenna, which in turn generates pulsed THz waves. The model successfully solves such Multiphysics problems by decoupling optical response from THz EM computations.

Furthermore, we find that graphene has little impact on incident laser pulse transmission and reflection. However, it enhances carrier dynamics. According to Fig. 2, graphene creates a more efficient electron transfer path and absorption than moving electrons through the SI-GaAs substrate due to loss in carrier recombination. The transient current is sharper in the time domain, as shown in Fig. 3, leading to a high-frequency enhancement. From Fig. 4, the radiation intensity at 1THz is enhanced by 4.2dB with Graphene.

## V. SUMMARY

In this work, a Multiphysics and Multi-scale modeling for PCA is realized, numerically analyzed and validated. Multiphysical simulation is achieved through effective de-coupling of semiconductor and optoelectronic responses and integrating subsequently to bridge THz and optics under a coherent platform. Graphene-enhanced PCA is simulated and validated as well with 4.2dB better performance than that of conventional bowtie PCA at 1THz.